\documentclass[twocolumn]{aastex631}
\usepackage[utf8]{inputenc}
\usepackage{comment}
\usepackage{amsmath}
\usepackage{enumerate}

\newcommand{\qso}{SDSS 0956+5128}
\newcommand{\ha}{H$\alpha$}
\newcommand{\hb}{H$\beta$}
\newcommand{\lya}{Ly$\alpha$}

\newcommand{\kms}{km s$^{-1}$}

\shortauthors{Rusakov et al.}

\graphicspath{{./}{}}

\begin{document}

\title{A broad-line quasar with unexplained extreme velocity offsets: post-shock outflow?}

\author[0000-0001-7633-3985]{Vadim Rusakov}
\affiliation{Niels Bohr Institute, University of Copenhagen, Lyngbyvej 2, K\o benhavn \O~2100, Denmark \href{mailto:rusakov124@gmail.com}{rusakov124@gmail.com}}
\affiliation{Cosmic Dawn Center (DAWN)}

\author[0000-0003-3780-6801]{Charles L. Steinhardt}
\affiliation{Cosmic Dawn Center (DAWN)}
\affiliation{Niels Bohr Institute, University of Copenhagen, Lyngbyvej 2, K\o benhavn \O~2100, Denmark}

\author[0000-0001-7825-0075]{Malte Schramm}
\affiliation{Graduate school of Science and Engineering, Saitama Univ., 255 Shimo-Okubo, Sakura-ku, Saitama City, Saitama 338-8570, JAPAN}

\author[0000-0002-9382-9832]{Andreas L. Faisst}
\affiliation{California institute of Technology, 1200 E California Blvd, Pasadena, CA 91125, USA}

\author[0000-0001-5382-6138]{Daniel Masters}
\affiliation{California institute of Technology, 1200 E California Blvd, Pasadena, CA 91125, USA}

\author[0000-0001-5846-4404]{Bahram Mobasher}
\affiliation{Department of Physics and Astronomy, University of California, Riverside, 900 University Avenue, Riverside, CA 92521, USA}

\author[0000-0001-8451-5316]{Petchara Pattarakijwanich}
\affiliation{Department of Physics, Faculty of Science, Mahidol University, 272 Rama IV Road, Ratchathewi, Bangkok, Thailand}

\begin{abstract}

The quasar SDSS 0956+5128 exhibits three distinct velocity components with large offsets in emission: the systemic velocity of [O {\sc ii}], [O {\sc iii}] and [Ne {\sc iii}] narrow lines have redshift $z=0.7142$; broad Mg {\sc ii} line is shifted by $-1200$ \kms{} with respect to the narrow lines; broad \ha{}, \hb{} lines are at $-4100$ \kms{}.  We present new Hubble Space Telescope spectra of \lya{} and C {\sc iv} emission lines and high-resolution images of the quasar.  The offsets of these lines are consistent with the velocity component of the Balmer emission, and the photometry in optical and near-infrared wavelengths does not show any signs of recent mergers in the host galaxy or irregularities in the location of the quasar.  The data do not confirm predictions of the previous most-likely hypotheses involving a special orientation and morphology of the quasar disk, such as in the recoiling black hole scenario, neither it is consistent with accretion disk winds.  Instead, based on the cumulative evidence, we propose a new scenario, in which the broad line region is in the state of outflow caused by a strong shock wave, with a supernova as a possible event for producing the shock ejecta.

\end{abstract}


\section{Introduction} \label{sec:intro}



Although there initially appeared to be several subtypes of active galactic nuclei (AGN) and quasars (QSO) \citep{antonucci1993}, it has been believed for approximately three decades that nearly all observed AGN are consistent with being similar objects observed from different lines of sight (\citealp{urry1995,miller1991,bailey1988,lawrence1982,antonucci1985}; other references in \citealp{antonucci1993}).  One consequence is that all quasar spectra exhibit the same set of significant spectral lines: (1) broad emission (or absorption) lines (BEL) including C {\sc iv}, \lya{}, H$\alpha$, H$\beta$ and Mg {\sc ii}, among others, with Doppler widths of $\sim 10^4$ \kms{} \citep{vanden_berk2001,murray1995,wills1993}, that are normally virialized \citep{shapovalova2001,dietrich1998,korista1995}; (2) narrow emission lines, including [O {\sc iii}]$\lambda\lambda$4959,5007, [Ne {\sc ii}]$\lambda$3869 and [O {\sc ii}]$\lambda$3727 with widths of $\sim 500-1000$ \kms{}; and (3) a wide range of possible narrow emission or absorption lines from the host galaxy, with widths of $<500$ \kms{}.  These observations have led to a physical picture in which this emission originates, respectively, from a broad-line region (BLR) $\sim$ 1 pc from the central supermassive black hole, a narrow-line region (NLR) $\sim 10^3$ pc away, and the remainder of the host galaxy, which might extend to $\sim 10^4-10^5$ pc out.  Although inflows or outflows might skew the profiles of these lines (cf. \citealp{strateva2003,eracleous1995,gaskell1982}), they all emanate from objects in the same host galaxy, and thus are likely centered at the same Doppler velocity (equivalently, redshift) with respect to the observer.

However, a single quasar among the more than $10^5$ quasars found in Sloan Digital Sky Survey (SDSS) appears to be entirely incompatible with this model.  SDSS J095632.49+512823.92 (hereafter, \qso{}) exhibits three distinct and non-typical features that in combination make this object unique \citep{steinhardt2012} (hereafter, S12).  

At first, there are three significantly different velocity components, corresponding to $z = 0.714$, $z = 0.707$, and $z=0.690$.  The narrow line emission of [O {\sc ii}], [O {\sc iii}] and [Ne {\sc iii}] provides the systemic redshift of the galaxy $z=0.714$.  The blueshift of broad Balmer emission lines and Mg {\sc ii} from the host galaxy is $\sim 4100$ and $\sim 1200$ \kms{}, respectively.  It is not surprising on its own to observe three components.  For example, such objects as I Zw 1 are known to exhibit more than two velocity systems with blueshifts from the systemic redshift \citep{Laor1997,vestergaard2001}.  However, unlike such quasars, the broad emission lines in \qso{} are not double-peaked or strongly skewed (as in the examples that were studied in detail in \citealp{eracleous1995,tsalmantza2011,eracleous2012}).  They are symmetric and completely offset and therefore appear to be consistent with some kind of physical offset of the BLR, such as an outflow.

Only a handful of objects studied in outflows have been observed with nearly as high offsets in \hb{}, \ha{} and Mg {\sc ii}.
In fact, offsets of the similar magnitude have been found more often in higher-ionization lines, such as Si {\sc iv}, C {\sc iv} or higher, and even so in the extreme tail of their offset distribution, as shown by \cite{yu2021} in SDSS DR7 \citep{shen2011} and other quasar samples.  The objects with symmetricaly-offset lines have been studied in connection with the recoiling black holes (BH; \citealp{bonning2007}).  In fact, no recoil candidates have been identified with the BLR velocity as high as in the \qso{} which is almost twice the second highest offset \citep{chiaberge2017}.  

In addition, the broad Mg {\sc ii} line, which is typically found to be consistent with the systemic velocity \citep{shen2016,hewett2010}, is offset by $-1200$ \kms{} from the host lines, and thus by $+2900$ \kms{} from H$\alpha$ and H$\beta$.  No other quasar spectra in the SDSS that have coverage of Balmer lines and Mg {\sc ii} are known to exhibit a strong velocity offset between these lines in the BLR, including the candidates for BH recoils (eg., \citealp{bonning2007}).  Indeed, such an offset between Mg {\sc ii} and hydrogen lines, which have similar ionization potentials, should not be possible if both are emitted from nearby parts of the BLR.  

It is worth noting that originally S12 reported that \ha{} and \hb{} broad lines are asymmetric, while Mg {\sc ii} is symmetric, contrary to the claim that all lines are symmetric in this work.  This is because there does not appear to be a strong asymmetry indicative of different velocity components in the Balmer lines.  It may rather be attributed to the clumpy nature of the BLR leading to a weak skew of the shapes of broad lines (\citealp{risaliti2005,risaliti2002}; references in \citealp{elitzur2006}).

Several mechanisms or explanations were considered in S12 to describe \qso{}, but none seems capable of producing all of the observed features.  The ideas included multiple objects along the line of sight, accretion disk winds, special morphological configurations or a recoiling black hole.  It is unclear which other known mechanism might be responsible for the behavior observed in \qso{}.  However, one way to distinguish between possible mechanisms is that different explanations make distinct predictions for the velocities of higher-ionization lines.

To examine the applicability of different scenarios, additional observations of high-ionization lines, like C {\sc iv}, are needed.  Combining it with the high-resolution photometry of the central region allows a test of these hypotheses.

Here, new {\it Hubble Space Telescope (HST)} observations of C {\sc iv} and \lya{} lines and photometry of the central region of the host galaxy are presented and analyzed.  Based on these data, we attempt to distinguish between the previously proposed mechanisms.  As none of them appears to match these observations, a new mechanism is proposed to explain \qso{}.

New {\it HST} measurements of the previously unobserved C {\sc iv} and \lya{} BELs in the UV spectrum (\S~\ref{sec:hstspec}) and photometric observations of the host galaxy at high resolution (\S~\ref{sec:hstphot}) are presented in the following sections.  The broad UV lines are symmetric and appear to have the offset consistent with the Balmer lines.  No irregularities are seen in the photometry of the central region of the galaxy.  While the QSO is seen as spatially offset, it is within $3\sigma$ resolution.  \S~\ref{sec:hypotheses} reviews hypotheses that were previously proposed to explain \qso{}, such as a combination of the recoiling BH scenario and a double-peaked emitter, and describes why it is a challenging problem.  Instead, the cumulative evidence shows a peculiar velocity pattern of the BLR consistent with a strong shock outflow from the central region.  Discussion of the observations and a new hypothesis along with final thoughts are presented in \S~\ref{sec:interpretation}.

This work adopts a flat $\Lambda$CDM cosmology with $\Omega_m = 0.3$, $\Omega_{\Lambda} = 0.7$, and $H_0 = 70~\mathrm{km~s}^{-1}~\mathrm{Mpc}^{-1}$ throughout.

\section{HST Spectrum} \label{sec:hstspec}

The near ultraviolet (NUV) spectrum of \qso{} was recorded with the Space Telescope Imaging Spectrograph (STIS) onboard the {\it HST} (Proposal 15872; \citealp{steinhardt2019}). The first order G230L long slit grating was used allowing for spatially resolved spectroscopy with $0.025''$/pix ($\sim 180$ pc/pix at $z=0.7142$) and low-to-medium spectral resolution with $R\sim500-1010$ (translates to $600-300$ \kms{} resolution at short to long wavelengths) in the NUV.  The {\it HST} pipeline CALSTIS \citep{sohn2019} used to reduce the spectrum allowed the detection of C {\sc iv} and \lya{} lines. The NUV spectrum allowed the detection of C {\sc iv} and \lya{} lines. 

The background in the spectrum was fitted with the exponential function in linear wavelength assuming that the continuum is dominated by the quasar emission.  The wavelength windows used for the fit ($1290 - 1460$, $1580 - 1810~\text{\AA}$) were expected to have little or no contamination from prominent line emission or Fe {\sc ii,iii} pseudo-continuum described in \cite{vestergaard2001} and were similar to the clear quasar-continuum windows in \cite{francis1991}.   The slope of the model continuum was used to validate the quasar nature of the spectrum.  Calculated as $F_{\nu1}/F_{\nu2} = (\nu_1/\nu_2)^{\beta}$, it was found to be $\beta = -1.18 \pm 0.01$, which loosely agrees with the observed distribution of slopes of typical quasars, although in the redder part \citep{davis2007}.  The model was used for the calculation as a cross-check, as the observed spectrum did not cover the wavelengths normally used above $1850\text{\AA}$.

\begin{figure*}
    \centering
    \includegraphics[width=0.8\textwidth]{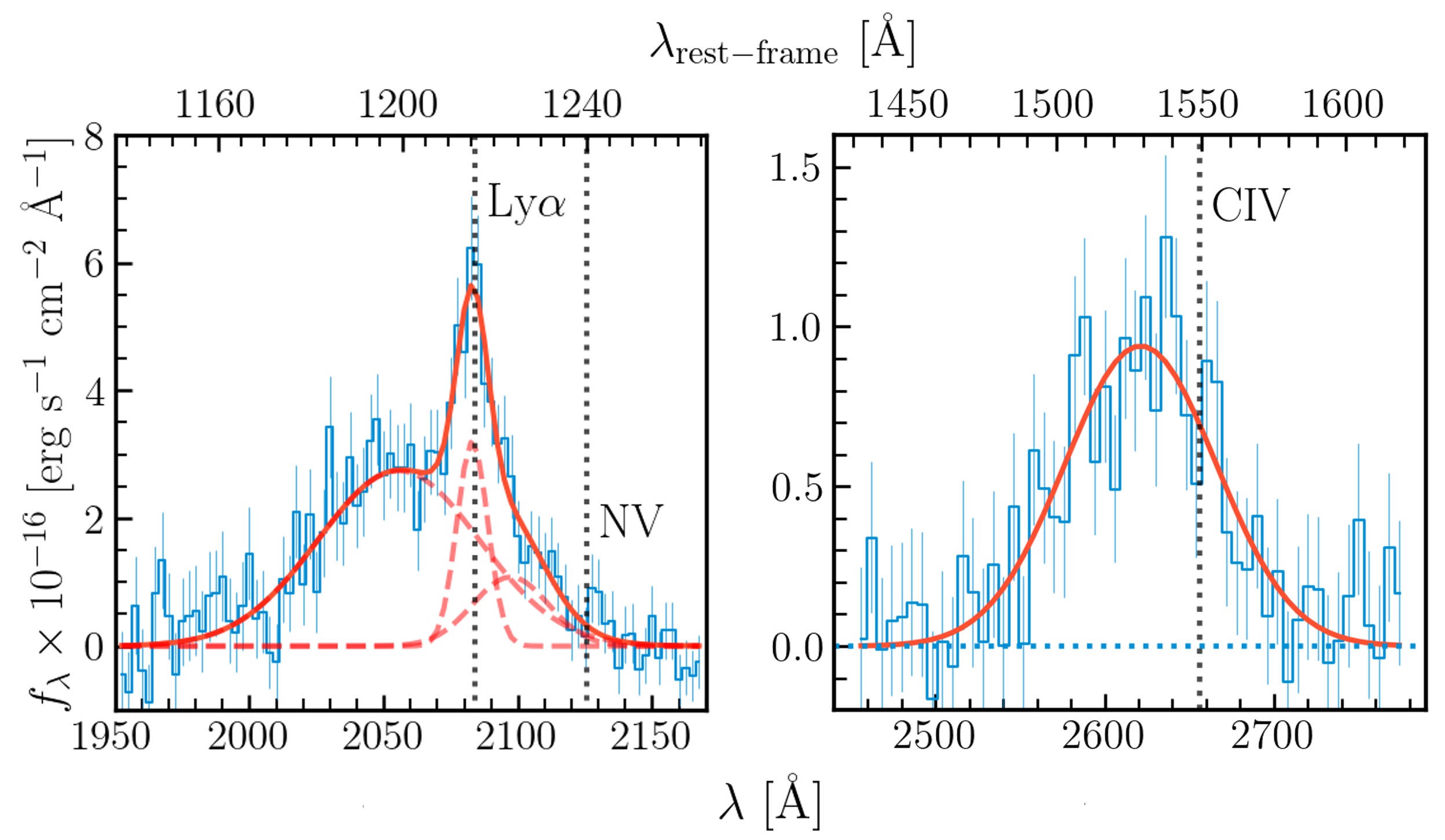}
    \caption{\lya{} $\lambda$1215 and C {\sc iv} $\lambda$1549 spectrum profiles. Vertical dotted lines indicate the systemic redshift of line centroids. \lya{} profile shows significant flux excess at $\sim 2055~\text{\AA}$ in the observed frame, fitted here as the offset broad \lya{} emission.  The excess of flux in the long-wavelength tail is suggested to arise from the offset N {\sc v} broad emission, as there is a hint at the narrow N {\sc v} $\lambda 1240$ line, which is unresolved in this spectrum.  The spectrum is resampled to $\sim 3~\text{\AA}$ per bin.  C {\sc iv} profile does not appear affected by strong emission or absorption features when resampled to $\sim 6~\text{\AA}$ per bin.  Verifying the BH mass using the C {\sc iv} estimator suggests the width of the fitted profile is the real C {\sc iv} width (see text).  The line appears completely offset from the host quasar redshift.}
    \label{fig:lines}
\end{figure*}

The background-subtracted C {\sc iv} and \lya{} are shown in Figure~\ref{fig:lines}. The systemic \lya{} line is best fit with a Gaussian component at $z = 0.7135 \pm 0.0003$, at $2.3\sigma$ deviation from the systemic redshift measured in S12.  The line profile is strongly asymmetric, with the most likely contribution from a blueshifted broad \lya{} component. This component is best fit at $z=0.6909 \pm 0.0026$ with $\mathrm{FWHM}\sim 11511 \pm 1781$ \kms{}. This profile is consistent with the observations of Balmer lines in S12, where the narrow peaks are accompanied by the completely blue-shifted broad emission. When considering other possible contributions to this blueshifted excess, Si {\sc iii} $\lambda$1206.5 and O {\sc v} $\lambda$1218 lines coincide with the profile at $z=0.707$ and $z=0.690$, respectively (associated with two offsets identified in S12).  However, neither the expected narrow width of these lines, nor the expected low flux can explain the strong broad excess that is best described by the broad \lya{} emission.

Additionally, there appears to be a broad excess of flux in the red wing of the profile, which could not be caused by either Si {\sc iii}, O {\sc v}, Fe {\sc ii} or Fe {\sc iii} lines at any one of the involved redshifts.  Including this component of the profile improves the fit.  However, the widths and positions of the broad \lya{} and the unknown component become less constrained, as reflected in their uncertainties (see Table~\ref{tab:spectrum}).  The small bump at rest wavelength $1240~\text{\AA}$ suggests that the unknown broad component may be the offset N {\sc v} emission.  With this, the fit of the whole profile was produced with $\chi^2_{\nu} = 0.77$, where the suspected broad N {\sc v} line had $\mathrm{FWHM}=2907 \pm 2862$ \kms{} at $z=0.6914 \pm 0.0051$.  There is however no distinct narrow N {\sc v} $\lambda$1240 component.  
The offset of the broad \lya{} and N {\sc v} components is at $1.4$ and $3.2\sigma$ levels, respectively.

The redshift of C {\sc iv} is measured at $z=0.6907 \pm 0.0008$ with $\chi^2_{\nu} = 0.69$ per degree of freedom. Although due to low spectral resolution and spectral purity the line appears to be strongly affected by the noise, it is above 4.2$\sigma$ noise level. The spectrum was resampled to improve its signal to noise ratio. No other distinct emission or absorption features could be observed; however, the line can usually be contaminated by some narrow features.

As the left wing of the profile at $1490-1520~\text{\AA}$ (rest frame) could be affected by the Fe {\sc ii} and Fe {\sc iii} and Si {\sc ii} at $1531~\text{\AA}$ emission at the systemic redshift \citep{vestergaard2001}, the width of the C {\sc iv} fit was cross-checked by estimating the BH mass.  The empirical mass estimator from \cite{dallabonta2020} for single epoch measurements yields $log(M_{BH}/M_{\odot}) = 8.80 \pm 0.16$, which is $1.0\sigma$ away from the Mg {\sc ii}-based estimate of $log(M_{BH}/M_{\odot}) = 8.65$ from S12.  The relations based on the C {\sc iv} line width are less well constrained than those made with the \hb{} line mainly due to the lack of observations in the UV.  The literature on the most commonly identified agreements and differences in the estimators has been summarized in \cite{dallabonta2020}.  Their empirical estimator was calibrated against identified correlations in the residuals with the reverberation mapping measurements.  In this specific case if the Fe emission acted to decrease the width of the C {\sc iv} profile, this would only make the disagreement with other estimators worse.  Therefore, the cross-check above is used as the main argument for the broad C {\sc iv} line producing dominant contribution to the profile.  It is concluded that the C {\sc iv} profile is observed as emission coming entirely from the BLR with the velocity offset.  However, it should be stated that applying mass estimators to strong outflows makes the estimates not trustworthy and speculative.

\begin{deluxetable*}{@{}lcccc@{}}
    \label{tab:spectrum}
    \tablecaption{Measurements of lines in NUV spectrum (this work), optical and NIR (S12). The offsets are stated with respect to the systemic redshift ($z=0.7142$) reported in S12.}
    \tablewidth{0pt}
    \tablehead{
    \colhead{Line} & \colhead{Component} & \colhead{Redshift} & \colhead{FWHM (\kms{})} & \colhead{Offset (\kms{})}
    }
    \startdata
        N {\sc v} $\lambda$1240? & BEL & 0.6914 $\pm$ 0.0051 & 4917 $\pm$ 4840	& -3996 $\pm$ 893 \\
        C {\sc iv} $\lambda$1549 & BEL & 0.6907 $\pm$ 0.0008 & 12491 $\pm$ 953 & -4107 $\pm$ 143 \\
        Ly$\alpha$ $\lambda$1215 & BEL & 0.6909 $\pm$ 0.0026 & 11511 $\pm$ 1781 & -4079 $\pm$ 457 \\
        \ha{} $\lambda$6563 & BEL & 0.690 & $\sim$7200 & -4100 \\
        \hb{} $\lambda$4861 & BEL & 0.690 & $\sim$7200 & -4100 \\
        Mg {\sc ii} $\lambda$2798 & BEL & 0.7071 $\pm$ 0.0006 & 12800 $\pm$ 490 & -1200 \\
        Ly$\alpha$ $\lambda$1215 & NEL & 0.7135 $\pm$ 0.0003 & 1978 $\pm$ 565 & -123 $\pm$ 48 \\
        {[}O {\sc iii}{]} $\lambda$4959 & NEL & 0.714 &  &  \\
        {[}O {\sc iii}{]} $\lambda$5007 & NEL & 0.714 &  &  \\
        {[}O {\sc ii}{]} $\lambda$3727 & NEL & 0.714 &  &  \\
        {[}Ne {\sc iii}{]} $\lambda$3881 & NEL & 0.714 &  &
    \enddata
\end{deluxetable*}

\section{HST Images} \label{sec:hstphot}

As part of the same proposal (15872), ACS/WFC camera of the {\it HST} was used to take 6 dithered images in F606W and 4 in F850LP filters to produce the combined images totaling 2075 and 2300 second exposures, respectively.  The individual exposures were aligned, background-subtracted and drizzled by \texttt{grizli} pipeline \citep{brammer2021}.  The final products represent the combined mosaics of these dithered images for each filter.

The reconstruction of the quasar 
emission was performed by using an effective PSF (ePSF) constructed with \texttt{photutils} package \citep{photutils} and based on 6 foreground stars in the images utilised here.
It is an empirical model based on the selection of stars in the images \citep{anderson2000,anderson2016}.  It is produced by simply measuring the flux of stellar sources in the vicinity of the target at individual pixels and it represents a map of fractional flux produced by a point source given the optics and the detector sensitivity, i.e. the instrumental PSF scaled by the pixel sensitivity map.  This method was shown to be more precise and numerically efficient than deconvolving the photometry and instrumental effects, assuming and then fitting an analytical function.  The advantages are particularly justified, when only a single point source is investigated.

\begin{figure*}
    \fig{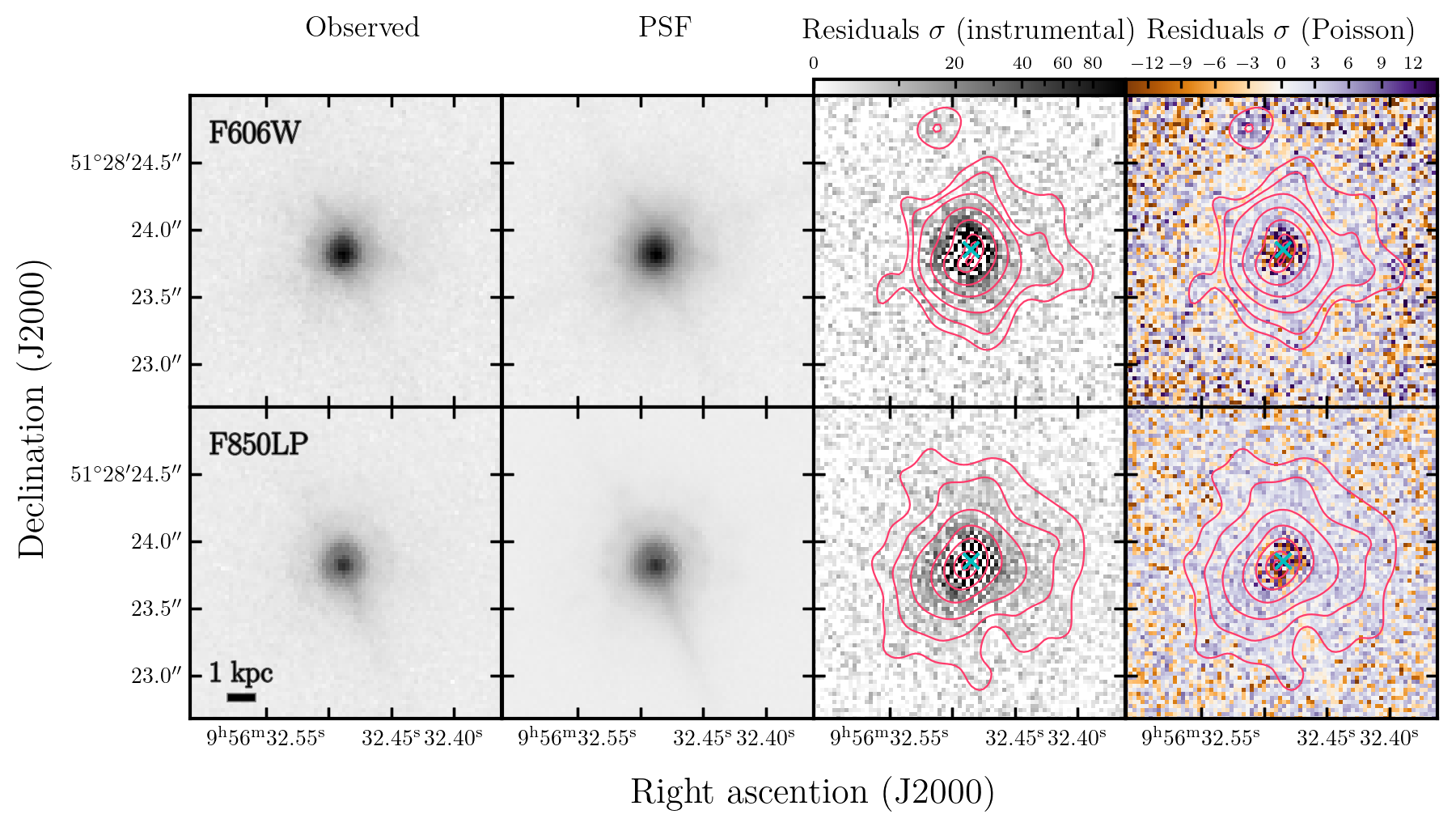}{\textwidth}{}
    \caption{HST/ACS images of \qso{}: observed image (first column); model PSF (second column); positive} residuals after subtracting the PSF in units of instrumental $\sigma$-noise (third column); residuals after subtracting the PSF in units of Poisson $\sigma$ (fourth column).  The top and bottom rows shows photometry in F606W and F850LP filters, respectively.  Contours show the residual host emission without strong eccentricity or possible post-merger disruption (smoothed with a 3.3-pixel Gaussian kernel).  The outer-most isophotes extend to the 50th percentile in F606W and F850LP indicating that the emission is more centrally concentrated in F606W and more extended in F850LP.  The best-fit centers of the quasar emission are marked with the cyan crosses, where the PSF centers are offset by 1.1 (F606W) and 1.4 (F850LP) pixels from the isophotal centers.  This offset is within the $3\sigma$ offset of $\sim 1.5$ pixels).  The scale is 0.03 arcsec per pixel (0.22 kpc).
    
    
    \label{fig:hstphot}
\end{figure*}

Figure~\ref{fig:hstphot} shows the residuals after fitting and subtracting the best-fit PSF from the images.  The fit was performed using the 2D image profile fitting code IMFIT \citep{erwin2015} with reduced $\chi_r^2=2.83$ for F606W and $\chi_r^2=1.36$ for F850LP images.  With the pixel scale of $0.03 \arcsec$ (0.22 kpc per pixel) the images allowed to resolve the host structure around the AGN.  The figure shows both the residuals in terms of the instrumental read-errors and the residuals in units of Poisson $\sigma$ defined using the average of the counts in the observed and model images.  These maps indicate that the residual host emission around the AGN is most significant in the F606W band (up to $5 \sigma$, Poisson), while the emission in F850LP (up to $2-3 \sigma$) is shallower and more extended outside of the central region than in F606W.  This is roughly in agreement with the relative amount of flux of the model SED emission of the host galaxy at the respective wavelengths presented in S12.

The residuals in either filter do not indicate any disruption or deformation in the central structure of the apparently elliptical light profile, which would be expected from a recent galaxy merger.  However, it was not possible to perform a detailed study of the host galaxy profile due to the overall shallow photometry.  The cyan `x'-marks indicate the best-fit centers of the quasar emission, which are misplaced from the center of the isophotal host emission by 1.1 (0.24 kpc) and 1.4 (0.31 kpc) pixels in F606W and F850LP, respectively.  Given the resolution, the offset up to $\sim1.5$ ($3\sigma$) pixels is allowed, which makes the best-fit QSO location consistent with the center of the host emission.



\section{Previously Proposed Explanations} \label{sec:hypotheses}

There exist several mechanisms that are responsible for creating different distinct velocity components in BLRs of quasars.  However, none of them appears to provide a full explanation for the observed features of \qso{}.  Those include outflows, quasar morphology and orientation or quasar motion with respect to its host that could correspond to one or both of the velocity components in \qso{}.  Specifically, S12 considered multiple objects along the line of sight, double-peaked emitter profile and a recoiling black hole.  Another considered mechanism is accretion disk winds that may be responsible for velocity profiles typically seen in most of the QSOs with outflows.  This section describes how the previous and new evidence from \qso{} fits within these scenarios and shows that none of them are capable of explaining the observation completely.

\subsection{Multiple objects along the line of sight}

Perhaps the simplest explanation for multiple velocity components would be multiple objects along the same line of sight.  However, this explanation cannot produce the broad lines observed in \qso{},  There is a unique set of narrow lines, consistent in their redshift, and two velocity components, represented by either broad Mg {\sc ii} or Balmer lines, but not both.  With two objects along the line of sight, two sets of narrow lines would have to be present, even if there was a specific combination with a strong Mg {\sc ii} broad emission and very weak Balmer lines in one object and a very weak Mg {\sc ii} and strong Balmer emission in the other.

In addition, the narrow lines argue against this explanation.  Narrow Balmer lines at the systemic velocity indicate that the central region of the presumed host galaxy is illuminated by the quasar.  The broad Balmer lines are bluer than the host.  Thus, if along the same line of sight, they would need to be in front of the host and unable to produce these features.  Therefore, a scenario with three distinct components in the same system has to be considered: emission of the host ($z=0.714$) in the NLR, Mg {\sc ii} ($z=0.707$) and Balmer ($z=0.690$) emission in the BLR.

\subsection{Disk winds}

In the context of a single QSO system, one common mechanism thought to be responsible for outflows in the BLR is accretion disk winds (e.g., models by \citealp{murray1995,mathews1977}).  In support of these models, \cite{gaskell1985} found evidence that BLR clouds could be radiatively accelerated.  It was proposed and modeled that radiation pressure could be responsible for accelerating the gas clouds radially outwards \citep{emmering1992,everett2005}.

In such a scenario, the velocity of the outflow is highest for the high-ionization material of the BLR and decreases for species with lower-ionization energies.  Examples of such profiles can be seen in \citet{meyer2019,shen2016,marziani2010,marziani1996,brotherton1994}.  As the ionization potentials are correlated with the radial distances of emission lines, the lines like
N {\sc v}, C {\sc iv} and other high-ionization lines experience some of the largest relative velocities reaching several thousands km per second \citep{yu2021}.  The intermediate- to low-ionization emission, including hydrogen lines and Mg {\sc ii} is typically seen accelerated to at most a few hundred \kms{}.  Often \ha{}, \hb{} and Mg {\sc ii} are consistent with the systemic velocity to within $\sim 200$ \kms{} \citep{shen2016,hewett2010}.

Although such a negative velocity gradient could produce a difference in emission line velocities, the difference seen between the hydrogen lines and Mg {\sc ii} in \qso{} is too steep and too high for the lines of such similar ionization potentials.  For instance, \citet{meyer2019} show that the offset velocities of Mg {\sc ii} and other low-ionization lines are typically identical in spectroscopically observed SDSS quasars, with velocity shifts well within a few hundred \kms{} at $z < 7$.  Moreover, in the rare cases where Mg {\sc ii} blueshifts of $\sim 10^3$ \kms{} are reported, they are generally connected with other mechanisms (as in the recoiling black hole candidate, 3C 186, in \citealp{chiaberge2017}).

The similarity between ionization potentials of \hb{} and Mg {\sc ii} alone may not be indicative of their physical proximity, especially if the lines originate from fully and partially-ionized regions and their emission is driven by different excitation mechanisms.  However, the correlation between the radius of the species in the BLR and the accompanying continuum luminosity (at $\lambda=5100\text{\AA}$ and $3000~\text{\AA}$) indicates that they are likely coming from the adjacent gas shells in the BLR.  Figure~\ref{fig:rl_relations} shows estimates with several of such empirical relations.  They may not be applicable to all quasars, because they are based on different sample selections.  However, any pair of radius estimates show that \hb{} and Mg {\sc ii} here are closely spaced and \hb{} is closer to the center on average.

Thus, if a negative velocity gradient is responsible for the extreme offsets of the hydrogen and Mg {\sc ii} lines in \qso{}, there should be an even larger difference when Mg{\sc ii} is compared with high-ionization lines.  For example, such trend was found in the offsets of emission lines of I Zw 1 \citep{Laor1997,vestergaard2001} and other quasars \citep{corbin1990,espey1989,wilkes1986}.  However, as shown in \S~\ref{sec:hstspec}, this is inconsistent with the new {\it HST} observations presented in this work, especially with the C {\sc iv}, possibly N {\sc v} and the hydrogen lines being symmetric and part of the same velocity system at $-4100$ \kms{}.


\subsection{QSO jets}

Alternatively, QSO jets that provide a way of losing the angular momentum for the supermassive black hole (SMBH) can cause QSO outflows \citep{Zheng1990}.  Even though the jets are highly collimated, they may cause outflows in the BLR assuming a high covering factor of the clouds.  It was shown in \cite{Zheng1990} that double-peaked Balmer emission can be produced in such AGN models.
However, this does not explain the symmetric single-peaked lines observed in \qso{}.




\subsection{Double-peaked emitter}

In another scenario, large line shifts of over 4000 \kms{} observed in some galaxies can be described by models with non-axisymmetric accretion disks \citep{strateva2003}: flattened, eccentric disks (or other forms of asymmetries) and a preferred inclination angle \citep{chen1989,eracleous1995}.  The problem in this case is that BELs in such systems produce double or asymmetric profiles.  S12 reported that the Balmer lines in \qso{} have asymmetric profiles, with a faint, broad component and that Mg {\sc ii}, in contrast, is symmetric.  However, the asymmetry appeared only in one of three independent observations (S12).

If the model of a double-peaked emitter is allowed to have sufficiently many parameters, it is possible to produce a reasonable fit to the spectrum in most cases.  However, in this case one double-peaked emitter is not able to explain scenarios in which different lines show different offsets.  
At least two eccentric emitter components are required to provide an explanation for the two offsets, but still not sufficient to produce a symmetric (non-double) Mg {\sc ii}, \lya{} and C {\sc iv} (and possibly N {\sc v}), unless these lines have double profiles with the second component being very weak.

\subsection{Recoiling black hole}

Finally, S12 suggested that \qso{} may be a recoiling BH in a post-merger galaxy, in which the BLR is a combination of eccentric and circular components.  A SMBH resulting from coalescence of two smaller SMBHs can receive a kick in some direction, depending on the rotation properties of the system \citep{campanelli2007b,schnittman2007,loeb2007}.  

It was shown that such a BH can still exhibit symmetric BELs with the offset of over 1000 \kms{} \citep{merritt2006,loeb2007}.  However, an offset of $\sim 4000$ \kms{} would be difficult to produce.  Some analytical and numerical considerations limit the maximum recoil velocity to $\sim 4000$ \kms{} \citep{baker2008,campanelli2007a,campanelli2007b} or even lower \citep{healey2014}, while others can reach up to $\sim 5000$ \kms{} for special configurations \citep{lousto2012}.  In observations, one potential candidate is the object CID-42 \citep{civano2010,civano2012,blecha2013}, with the offset of up to 1300 \kms{} detected in the broad Balmer emission.  Another candidate for a SMBH recoil with more BEL detections, 3C 186 \cite{chiaberge2017}, shows the highest known offset of $\sim 2100$ \kms{}, which is second only to \qso{} with the offset nearly twice as high ($\sim 4100$ \kms{}).  In addition, the spectroscopic velocity profile of 3C 186 appears to be constant with respect to the ionization potential of C {\sc iv}, C {\sc iii}] and \lya{}, as well as Mg {\sc ii}.

In \qso{}, C {\sc iv} (and possibly N {\sc v}) and the hydrogen lines are consistent with this scenario; however, Mg {\sc ii} is not.  Given the similarity of the ionization potentials of hydrogen and Mg {\sc ii} lines, it seems very unlikely that the Mg {\sc ii} region could be moving $\sim 2900$ \kms{} slower than the rest of the recoiling BLR.  One possibility is to assume that Mg {\sc ii} is consistent with the recoil scenario and all of the BLR is offset by at least the offset of Mg {\sc ii} ($-1200$ \kms{}).  Then even slightly higher-ionization lines, including the hydrogen lines, could arise due to an eccentric emitter disk that introduces additional offset on top of the recoil and causes asymmetry of the Balmer lines.  Therefore, in this scenario, the Balmer and Mg {\sc ii} emission must come from separate locations, with the latter being further out and the higher-ionization lines, like C {\sc iv} or N {\sc v}, must also have the same asymmetry, as reported for the Balmer lines in S12.

In support of the idea of the recoiling BH, S12 showed using ground-based photometry that the peak of intensity in \qso{} was possibly offset from the center of light in the host galaxy, although the data had low resolution and systematic effects in PSF fitting were not ruled out.  Additionally, photometric decomposition showed that the host is a dusty galaxy, which can be consistent with various states of galaxy evolution, including a possible post-merger.  S12 showed that based on the timescale for the BLR to sustain emission after the quasar accretion was disrupted, the recoil could have occurred in the past 140 Myr.  This could allow for the host galaxy to preserve the evidence of a recent merger on a less than dynamical time, detectable in the follow-up resolved observations.

However, the new evidence presented from {\it HST} does not hold any indications of a BH recoil that could result from a previous galaxy merger.  It is shown in \S~\ref{sec:hstphot} that the QSO, fitted with a PSF, is spatially consistent with location of the isophotal center of the residual galaxy emission and no strong irregularities in the structure are seen that could result from a recent galaxy merger, although the photometry is not deep enough to accurately fit the host morphology.  It remains possible that the strong signatures of the merger could have been erased to the point of being undetected at the given resolution.  Nevertheless, new observations of the high-ionization line C {\sc iv} and low-ionization \lya{} appear to be consistent with the offset of the Balmer lines, they are symmetric (see Section~\ref{sec:hstspec}).  The symmetry is contrary to the expectation from the BH recoil hypothesis.  It was also shown by S12 that the system lies on the standard $M_{BH} - M_{bulge}$ galaxy mass relation, with the virial mass of the SMBH of $log(M_{BH}/M_{\odot}) = 8.65$ (S12) lying close to the empirical relation in the plane of the galaxy luminosity and the black hole mass.  Therefore, this does not provide any sharp indications that the SMBH does not belong to the host.  In this context, the evidence suggests that the offsets in \qso{} do not appear to be caused by the motion of the quasar, but are rather intrinsic to the QSO.

In \S~\ref{sec:interpretation}, we argue that the radial velocity profile in the BLR of \qso{} is reminiscent of the post-shock outflow and use the idea of star formation in the accretion disks (eg., from \citealp{goodman2004}) to support this mechanism for causing strong outflows in the BLR of quasars.

\begin{figure}
    \centering
    \includegraphics[width=0.49\textwidth]{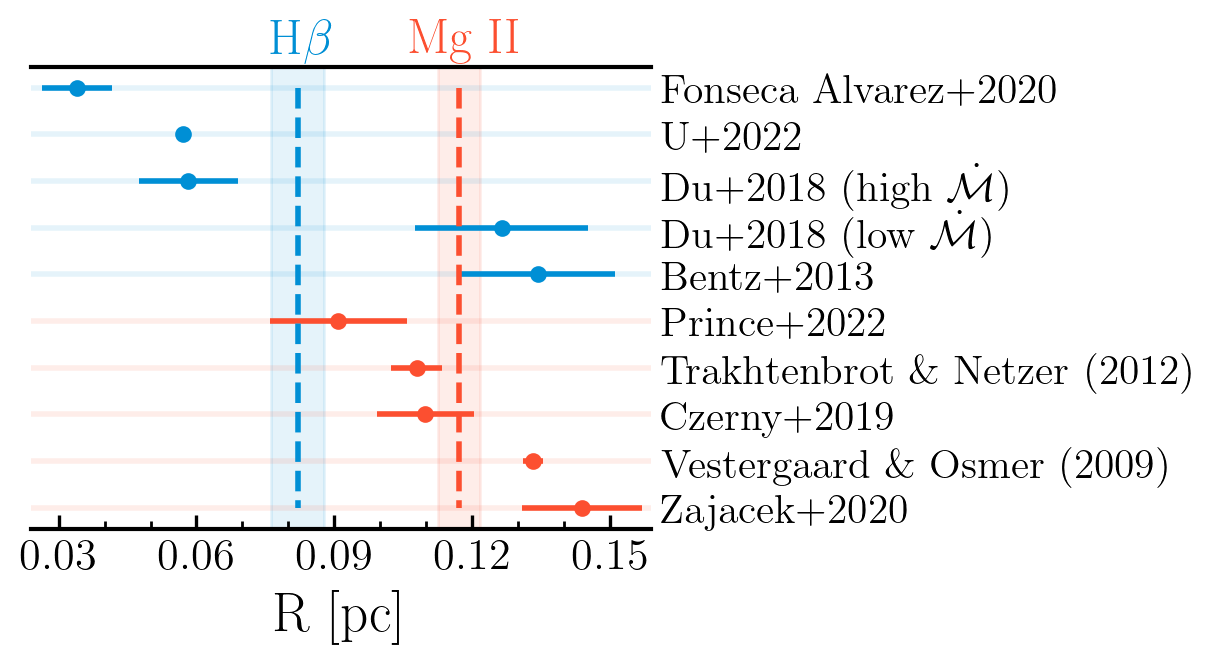}
    \caption{Compilation of various calculations of the radii of \hb{} and Mg {\sc ii} obtained using the respective luminosities $\lambda L_{5100}$ and $\lambda L_{3000}$ of the continuum in \qso{}.  Individual estimates are shown with 1-$\sigma$ uncertainties.  The average radii are shown as vertical dashed lines with their corresponding shaded uncertainty bars.  The estimates are based on the empirical radius-luminosity relations for quasars.  Some of the studies use very different sample selections, such as different accretion rates $\dot{\mathcal{M}}$ in \cite{du2018}, to define the relations.  Nevertheless, they produce similar radii for the two species here, where \hb{} is on average is closer to the center than Mg {\sc ii}.}
    \label{fig:rl_relations}
\end{figure}

\section{Interpretation} \label{sec:interpretation}

The observations of \qso{} appear to be inconsistent with known physical mechanisms and suggest that an outflow with very distinct velocity signature is responsible.

The event that produced the outflow must have an origin at the BH or close to it.  This appears to be the case, because the symmetry of the BELs suggests a spherically uniform outflow.  Additionally, the whole BLR appears to be affected, where the radial velocity is constant across a large fraction of the BLR, between the high-ionization region and low-ionization hydrogen lines.  Finally, there appears to be a drop-off of the outflow velocity starting from the location where Mg {\sc ii} is produced.  Therefore, it may be safe to assume any origin of such event within the inner boundary of the BLR, with its effects limited to the BLR outskirts.

Clearly, some extreme physical conditions must be at play.  The event has to be consistent with the outflow energetically and with the observed radial velocity profile, as indicated by the offsets of the broad lines and their expected radial distances.  The upper limit on the time of the event can be placed by using the velocity of the outflow.  

Below, a shock wave from an energetic explosion is postulated as a mechanism for causing an outflow.  It is shown that this mechanism can be consistent with the observed features in \qso{}.  Then the plausibility of the extreme physical conditions due to a supernova explosion is considered, along with a discussion of whether stars can exist in close proximity of a SMBH and how they can appear there in the first place.

\subsection{Post-shock outflow} \label{ssec:post_shock}

The velocity profile of the outflow as a function of radius in the BLR of \qso{} is strongly reminiscent of that of a ``shocked'' material \citep{taylor1950}, as it shown in this section.  It is expected that the radial profile in terms of increasing radial distances is as follows: C {\sc iv}, \lya{}, \ha{}, \hb{}, Mg {\sc ii}, where Mg {\sc ii} must be at least as far out as the Balmer line emission.  Based on the spectroscopic observations made in Section~\ref{sec:hstspec} here and in S12, the outflow velocity profile is constant at $\Delta V \approx -4100$ \kms{} starting at the location of the C {\sc iv} line and extending out to the hydrogen lines.  Then the radial velocity drops to $-1200$ \kms{} at the location of Mg {\sc ii}.  Assuming the velocities estimated from several spectra across an extended timeline are accurate, there would be only one explanation for different offset velocities of the Balmer lines and Mg {\sc ii}: the latter line must originate at larger radii due to its lower ionization energy.  In this case, the overall velocity profile is characteristic of a post-shock outflow, such as the outflows in the interstellar medium (ISM) produced by supernova (SN) explosions.

Assuming a uniform density of the medium, the evolution of shocks can be described as a three-phase process that starts with a {\it free expansion} at high pressure relative to the surroundings, during which the mass of the medium swept up by the shock wave along the direction of travel is insignificant compared to the mass of the shock material itself. Thus, the energy and momentum of the wave are conserved and the kinetic energy $E_k$ is: 

\begin{equation}
    E_k = M_{ej} V_{ej}^2 / 2,
\end{equation}

where $M_{ej}$ and $V_{ej}$ are the mass and velocity of the ejecta.  

As the shock shell expands in radius, it is opposed by a larger mass of the ambient medium.  When the two masses become equal, the shock enters the {\it Sedov-Taylor phase} \citep{taylor1950}.  There, the ejecta start to lose its momentum, while the matter is too heated to irradiate, so the system remains adiabatic. At this stage, the expansion of the shock radius $R_S$ with time $t$ is found to depend entirely on the initial kinetic energy $E_k$ and density of the medium $\rho_0$: 

\begin{equation}
    R_s(t) \propto E_k^{1/5} \rho_0^{-1/5} t^{2/5}.
\end{equation}

Finally, as the temperature of the shock drops, C, N, O ions start to recombine helping to cool the shock material efficiently, which starts to loose its energy radiatively until the flow becomes subsonic and merges with the surrounding medium.

\begin{figure*}
    \includegraphics[width=1.0\textwidth]{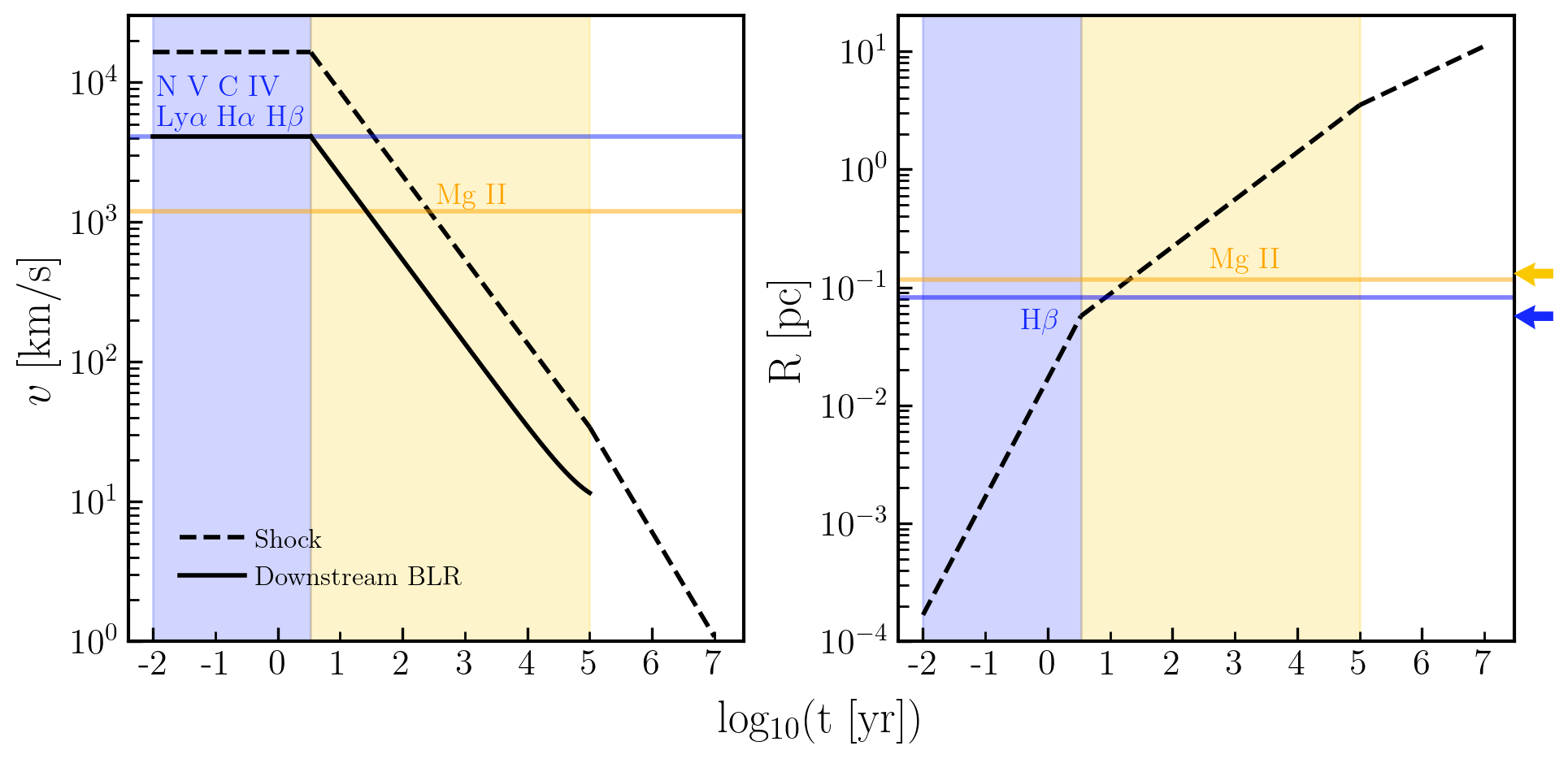}
    \caption{Velocity (left) and distance (right) profiles of a shock wave produced by an explosion of a star associated with the ejecta mass $M_{ej} = 2 \times 10^4 \mathrm{M}_{\odot}$ and kinetic energy $E_K = 5 \times 10^{55}$ erg. The BLR medium has a uniform density ($n_e=10^9$ cm$^{-3}$); other main assumptions are stated in the text. Shadowed regions represent different stages of a shock wave traveling through the medium: {\it free expansion} (momentum and energy conserving); {\it Sedov-Taylor phase} (adiabatic; energy conserving); {\it snowplow phase} (radiative and momentum conserving). {\it Left panel}: two observed offset velocities of BELs are indicated by the horizontal solid lines and correspond to different stages by color, as deduced from their predicted downstream flow velocities (black solid line). The predicted downstream velocity profile is calculated only for the first two phases, before the shock becomes nearly sonic and valid approximations break.  {\it Right panel}: predicted radial profile of the shock wave (black dashed line); the observed radii of \hb{} and Mg {\sc ii} are obtained using the mean of the estimates from the RL relations and shown by the horizontal lines with the same colors as before};  the yellow arrow indicates the Mg {\sc ii} radius predicted from the time when the shock reaches Mg {\sc ii} location, i.e. the time when the downstream flow coincides with the observed velocity of Mg {\sc ii} in the left panel; the blue arrow show the location of the boundary between the first two phases of the shock flow.
    \label{fig:sn-shock}
\end{figure*}

This evolution can be matched with the velocity components observed in \qso{}, as illustrated by the velocity and distance profiles of a shock wave in Figure~\ref{fig:sn-shock}.  In this interpretation, the velocities presented here sample the first two phases: C {\sc iv}, \lya{}, \hb{} and \ha{} represent the region in the {\it free expansion phase}, while Mg {\sc ii} is in the {\it Sedov-Taylor region}.  Also, in the analysis above the high-ionization N {\sc v} line\footnote{N {\sc v} is the line with the highest ionization potential in our sample, which places it closest to the SMBH.} was possibly identified with the offset placing it consistently with the lines in the first phase.  This phase is sampled well with 4 or 5 lines probably spanning a significant fraction of the BLR radial profile. However, the second phase is only seen with one BEL (Mg {\sc ii}). Fortunately, there have been two observations of this line 7 years apart to confirm the detection.

The velocity profile of the shock stages shown in Figure~\ref{fig:sn-shock} is annotated with the offset velocities of the corresponding broad lines.  While the source energy sets the initial velocity of the flow, the density of the BLR sets the distance and time scale of the shock profile.  Here, it was assumed that the density is uniform throughout the region and is equal to the lower limit of $n_e = 10^{9}$ cm$^{-3}$ in the BLR \citep{osterbrock1989,kwan1981}.  Higher densities act to shrink the distance and time scales.

The velocity that the ambient BLR material gains when it crosses over the shock boundary depends on the ratio between the speed of sound and the shock. The speed of sound in the BLR gas is $c_s \approx 10$ \kms{}, assuming the BLR is in photoionization equilibrium and has the uniform temperature of $T = 10^4$ K, which is the minimum required for photoionization \citep{osterbrock1989}.  Therefore, the outflows observed in \qso{} ($\sim 4100$ and $1200$ \kms{}) are strongly supersonic.  Hence, the shock wave should be expected even more so, such that the limit as the Mach number $\mathcal{M} \to \infty$ for $\mathcal{M} = v/c_s$ can be safely used.  In the reference frame with the shock at rest, the velocity of the post-shock material ($v_2$, downstream) can be related to the velocity of the pre-shock material ($v_1$, upstream) using the Rankine-Hugoniot jump condition:

\begin{equation}
    \frac{v_1}{v_2} = \frac{(\gamma+1) \mathcal{M}^2}{(\gamma + 1) + (\gamma - 1)(\mathcal{M}^2 - 1)},
\end{equation}

where $\gamma = 5/3$ is the adiabatic index for an ideal monoatomic gas.  The assumption here is that the flow is adiabatic and the entropy is constant across the shock boundary, which holds for the first two stages of the shock wave, before it becomes subsonic and its viscosity cannot be neglected leading to the start of radiative cooling.  At $\mathcal{M} \to \infty$, $v_2 \to 0.25 v_1$.  Therefore, we would expect the shock to have traversed the locations of the observed broad emission at $v_1 \approx 16400$ \kms{} and $v_1 \approx 4800$ \kms{} to produce the high ($v_2\sim4100$ \kms{}) and low ($v_2\sim1200$ \kms{}) observed velocity components in the BLR, respectively.

Interestingly, under the assumption of the BLR gas density of $n_e = 10^{9}$ cm$^{-3}$ two key model predictions are in agreement with the independent expectations from the empirical radius-luminosity (RL) relations.

First, the model correctly predicts the location of the Mg {\sc ii} gas.  This prediction is made by using the time when the calculated downstream flow (black solid line; left panel in Fig.~\ref{fig:sn-shock}) matches the observed velocity offset of Mg {\sc ii} (orange solid line): $t \approx 27$ yr.  Based on this time the predicted radius of the Mg {\sc ii} line is $0.131$ pc (blue arrow; right panel).  This closely agrees with the mean of the estimates from various RL relations: $R_{Mg {\sc II}} = 0.117 \pm 0.005$ pc (see Figure~\ref{fig:rl_relations}; \citealp{prince2022,zajacek2020,czerny2019,trakhtenbrot2012,vestergaard2009}).

Second, the location of the boundary between the first two phases ($\approx 0.057$ pc) agrees with the location of \hb{} from RL relations (blue solid line; right panel).  The average of several RL estimators yields $R_{H{\beta}} = 0.082 \pm 0.006$ pc (see Figure~\ref{fig:rl_relations}; \citealp{u2022,fonsecaalvarez2020,du2018,bentz2013}).  This is consistent with the observation that the hydrogen lines are part of the high velocity component.  These two predictions are possible only for a narrow range of BLR gas densities around $n_e = 10^{9}$ cm$^{-3}$ and for a constant density as a function of radius, where the time scale goes as $t \propto n_e^{-1/3}$ in the first phase and $t \propto n_e^{-1/5}$ in the second.


Predicting the mass and type of the candidate supernova star is beyond the scope of this work, but it is possible to estimate an order of magnitude mass of the ejecta and their kinetic energy.  The ejecta mass can be obtained by locating the boundary between the first two shock regions.  This radial distance encloses the mass of the swept up gas comparable to the mass of the shocked ejecta, as this is the approximate condition for termination of the first phase.

The boundary must lie between the hydrogen lines (eg., \hb{}) and Mg {\sc ii}, which are close spatially (see Figure~\ref{fig:rl_relations}), but correspond to different velocity components in the BLR.  Assuming \hb{} is at the boundary (as shown in the right panel of Figure~\ref{fig:sn-shock}), we can use the estimated radius of the boundary of $0.057$ pc and the BLR density of $n_e = 10^9$ cm$^{-3}$ to estimate the ejecta mass.
This produces $M_{ej} \sim 2 \times 10^4$ $\mathrm{M}_\odot$.  

With the velocities of the BEL offsets in the first phase, the kinetic energy of the event is $E_K = 5 \times 10^{55}$ erg\footnote{This energy is $10^4$ times higher than the kinetic energy of the ejecta of a core-collapse SN explosion of an $8~ \mathrm{M}_{\odot}$ star.}.  Here, it is assumed that most of this mass is contained in a thin shell at the radius of \hb{}.  Therefore, the exploded star must have had an initial mass $M_* \geq M_{ej}$.  Such objects have not been observed before, although theoretical considerations of instability growth in the accretion disks of quasars suggest a possibility for formation of stars with masses at least as high as $10^2 - 10^7~\mathrm{M}_{\odot}$ \citep{goodman2004}.  



\subsection{Origin of stars in AGN}

If a stellar explosion produced the observed features in the spectrum of \qso{}, how could a star appear in the vicinity of a SMBH in the first place?  And how could it be nearly as massive as $2 \times 10^4~M_{\odot}$?  The star might be formed either in the proximity of the SMBH or in a host galaxy and then captured into the accretion disk.  One way or another, it probably owes its large mass to the abundance of material in the accretion disk of a SMBH.

The growth of supermassive stars may stem from the capture of stars from the host galaxy.  The closest galactic source of stars is the bulge that encircles the quasar.  Stochastic encounters can result in significant changes of stellar velocities on a relaxation timescale that could lead to stars being deflected off their orbits.  A ``lucky'' star brought to the central object with the mass $10^{8.6}$ M$_{\odot}$ would not risk being disrupted due to tidal forces, unless it sinks into the black hole.  The reason is the tidal radius of BHs with masses $> 10^8$ M$_{\odot}$ becomes smaller than the Schwarzschild radius.  In fact, sinking into the BH would be a more likely outcome, as otherwise reaching an exact stable orbit around the BH would require acquiring a very specific velocity making it very unlikely.

Following on the idea of disk-star interactions as a potential mechanism for removal of angular momentum from AGNs \citep{ostriker1983}, \cite{syer1991} investigated a possibility for stars to be captured by the accretion disk for favorable inclinations of stellar orbits.  This mechanism was characterised by shorter than the relaxation timescale for delivering stars into the disk.  It could be possible as the growth of the gaseous disk and clouds would be expected to extend as far as to the edge of influence of the BH for the stellar bulge.
Stars would then be dragged down by the viscous forces to the stable circular orbit over time.  If such stellar migration is real, then according to the predictions of \cite{artymowicz1993}, after the QSO acquires a disk and turns on, in the span of $10^8$ yr there can be as many as $10^4$ captured stars. They would be dragged down the disk, enmassed by the disk material to as high as $\sim 10^2~M_{\odot}$ and evolve on their respective main-sequence timescale leading to Type II SNe.  For the case of \qso{} a single core collapse supernova with such initial mass would not be able to explain the observed BLR offsets.  Although $\sim 10^2-10^3$ such SNe simultaneously would, this could hardly be a plausible scenario.  In addition, it was argued that the stars might more likely be destroyed by the release of energy for circularization of their orbit in the accretion disk, as a very limited set of conditions for entry must be met to stay bound \citep{goodman2004}.

On the other hand, \cite{goodman2004} considered in situ star formation and growth of massive stars.  The main argument that stars can and possibly should form in BH accretion disks was that starting from a distance of $\sim 10^3~R_S$, where $R_S$ is the Schwarzschild radius, the disks are dominated by self-gravity and are expected to fragment via local dynamical instability.  \cite{goodman2004} considered fragmentation at the exact boundary, where the radiation pressure equals the gas pressure, as at this location, any fragments would be well-separated and therefore likely to grow via accretion of the surrounding gas.  The authors showed that given the strong accretion in the disk, the star is more likely to grow than fragment.  

The properties of the protostar would be defined by the instability.  As mentioned previously, the initial overdensity masses could reach anywhere in the range $10^2 - 10^7 M_{\odot}$ and are likely to grow by further accretion.  The mass growth continues until the gap roughly of the size of the protostar's Roche lobe is cleared in the accretion disk.  Finally, the material in the Roche lobe would contract until the star reaches the main sequence.  Such stars are likely to be supported by radiation pressure and experience significant mass loss, although the levels are not known for the very high masses over $10^2~M_{\odot}$.

In the theoretical framework of \cite{goodman2004}, the mass of a protostar in a disk of $10^{8.6}~M_{\odot}$ SMBH is expected to be $\sim 10^5 M_{\odot}$, if it forms at the edge of the low self-gravity of the disk.  The likely outcome of the SNe of stars with mass $> 10^2~M_{\odot}$ is a complete collapse to a BH \citep{fryer2001}, although an instability from $e^+e^-$ pair production may occur that would convert all of the stellar mass to ejecta at explosion.  Therefore the scenario in which a star with an initial mass $10^5~M_{\odot}$ evolves into a less massive $\sim 10^4~M_{\odot}$ star by mass loss and causes a pair-instability supernova that ejects all of the material in an explosion would be consistent with the observation in \qso{}.

Following \cite{goodman2004}, if the star is stable on the main sequence, it can be expected to migrate inwards, preserving the disk gap, and reach the tidal radius on the time ($\sim 10^5$ yr) shorter than the main sequence time ($\sim 10^6$ yr).  It would be of particular interest here to know whether the star is expected to disintegrate once on the main sequence or if it survives for some parameters.  Future numerical simulations of such processes would be of particular interest to answer these questions.


If star formation in accretion disks is common, how frequently would an \qso{}-like event occur?  \citet{goodman2004} estimate that if star formation in a QSO happens at a rate of one per viscous time, then given the approximate number density of QSOs at redshifts $0.5 < z < 1.5$, $10^{-5}$ massive stars per QSO per year should form, for a total of a few stars per year in a survey the size of SDSS.  

However, most of those stars will not produce an observable supernova.  \citet{goodman2004} argued that sinking into the SMBH is the typical outcome.  Fresh, hydrogen-rich material from the accretion disk can mix into the convective stellar core, so that the main sequence lifetime becomes longer than the timescale for reaching the event horizon.  Thus, their deaths would not be observable.

Perhaps if the accretion disk becomes sparse towards either near the Eddington luminosity or towards the end of a quasar's lifetime, there might be just a high enough density to form a massive star, yet not enough additional material to sustain that star until it reaches the event horizon.  The resulting supernovae might even provide a turnoff mechanism, so that quasars die with a bang rather than with a whimper.  As \qso{} lies at $z =0.7$, below which relatively few quasars are observed, it is likely to turn off in the near future.  It is therefore plausible to associate the extreme features of \qso{} with a short-lived turnoff mechanism.  If such a process lasts for $\sim 10^3-10^4$ yr (based on the timeline of the shock wave in Fig.~\ref{fig:sn-shock}), finding one example in all of SDSS is plausible.  

\section{Discussion}

It is argued here that none of the common and less so mechanisms, except for the shock wave, are likely to produce a spectroscopic signature observed in the BLR of \qso{}.  The scenarios that involve special morphology, alignment of multiple objects, radiation output of the QSO or SMBH mergers each have at least one sharp inconsistency with the evidence (see Section~\ref{sec:hypotheses}).  Instead, this work proposes the idea of shock waves to explain the observed spectroscopic offsets.  The toy model requires the tuning of only the BLR density parameter to match the velocity offsets in \qso{}, as well as the location of Mg {\sc ii} line.

To support or reject the idea of the supernova shock causing the outflow in the BLR it is necessary to collect more observational evidence.  Below is the list with a couple of possibilities.

\begin{itemize}
    \item {\it Other offset lines?} The shock mechanism predicts that there must be a continuum of velocity components in the second phase of the shocked BLR (see Section~\ref{ssec:post_shock} and Figure~\ref{fig:sn-shock}), as opposed to only one that was observed.  Although the most prominent BEL in the second phase (Mg {\sc ii}) was already observed, lower-ionization lines in the NUV, like O {\sc i}, Si {\sc ii} or C {\sc ii} are not detected in the HST spectrum here.  Instead, they could be targeted with higher signal-to-noise spectra.  Additionally, Fe K$\alpha$ line in the X-rays would be expected to have the high velocity offset of $4100$ \kms{}, as the other lines of that velocity component.  Probably, the direction of the offset and the line symmetry would depend on the position relative to the potential SN explosion site in the accretion disk.  If any of these lines are inconsistent with the prediction, this will rule out the hypothesis.

    \item {\it Traces of neutrinos and electromagnetic emission.} From another perspective, SN shocks act as neutrino production sites, through pion decays as a result of proton-proton interactions.  It was suggested that the diffuse neutrino background may be substantially contributed to by AGN stellar explosions \citep{zhu2021}.  In order to detect these events, it was shown (though for much less massive stars) that short bursts in the diffuse neutrino background and the ensuing lasting electromagnetic emission can be used.  Therefore, such signature may be used to search for potential candidates for follow-up observations or for testing models of neutrino and electromagnetic emission produced by energetic stellar explosions.

\end{itemize}

After all, the shock hypothesis here is an analytical solution based on the simple assumptions of shock-medium interaction and lacks a more detailed insight.  Therefore, numerical modeling of the involved processes would be required to test the idea more robustly.  Below are the suggestions for possible experiments and predictions.

\begin{itemize}
    \item {\it Downstream velocity as function of time.} 
    The BLR gas with different offsets starts mixing in the {\it snowplow} phase altering the velocity of the BLR outflow.  If one can calculate the outflow velocity of every one of the observed emission lines as a function of time, it will be possible to predict the blueshift that the lines should have at a time of a possible future observation.  Besides, it will also be possible to predict how the strength of the broad line emission will change over time, as the emitting gas moves away from its ionization zone.  Additionally, the density of the BLR assumed here corresponds to a rather conservative lower limit and it has a uniform radial profile, which can be tested by modeling the broad line emission observed in \qso{}.

    \item {\it Numerical simulation of stars in a strong gravitational potential.} One of the main problems in the shock hypothesis is what caused it.  The only possibility to create a shock wave of the required energy known to us is a potential supernova.  If the outflow profile in \qso{} is a signature of an SN (see Section~\ref{sec:interpretation}), then the mass of the ejecta is estimated to be $\sim 10^4~M_{\odot}$ and the kinetic energy is $E_k \sim 10^{55}$ erg. It is unclear whether stars can be ``safely'' delivered to accretion disks of SMBHs and then grow to such extent by accretion, but there exist theoretical models that predict formation of such supermassive stars in these extreme environments in the first place.  To shed more light on the possibility of the supermassive stars leading to SNe shocks, numerical simulations of their formation and evolution are required.  In this regard, the most important questions are: (1) whether a stellar explosion is close to being spherical in a strong gravitational field (to offset a BLR entirely rather than in the plane of the disk); (2) whether a proto-star can remain stable to reach and stay on the stellar main sequence; (3) if the star's main sequence lifetime is shorter than the timescale for spiralling down into the SMBH; (4) if the ``spiralling down'' timescale wins, whether it may be reversed in special cases when the QSO shuts down accretion and turns off.

\end{itemize}

Solving the problem posed by \qso{} may shed new light on the evolution of quasars and formation of massive stars in extreme environments.  There is a potential for studying the physics of star formation and evolution in QSO disks, as potentially a few stars per year may form in an SDSS-sized survey of quasars. In addition, if stellar explosions are shown to be possible, this may improve the understanding of extreme stellar evolution and BLR outflows.  However, it is less clear how frequent supernovae may be.  If the offset emission of the BLR in \qso{} is the result of an SN, it is likely only one such event in the SDSS, as no similar candidates have been reported.  Although, new multi-line observations of the evolved quasars with strong outflows may reveal more of these objects in the future.  On the other hand, it is believed that they are not observable and more likely to cross the event horizon before they explode, because their lifetimes are prolonged by the freshly accreted gas.  Although, this may not be the case in sparse accretion disks of the old quasars, where little gas may not be enough to support the stars before sinking into the SMBH.  Therefore, if the occurrence rate is low because the SN-induced velocity offsets of BLRs in old quasars are transient on some timescale, then it is possible that SNe are more frequent and could also play an important role in turning off QSOs.


\vspace{0.5cm}
We would like to greatly thank Gabriel Brammer for help with the data processing, Martin Pessah, Darach Watson, Marianne Vestergaard, Johan Fynbo, Lise Christensen, Jeremy Goodman, Adam Jermyn, Ersilia Guarini and Kasper Elm Heintz for helpful and insightful discussions.  The Cosmic Dawn Center (DAWN) is funded by the Danish National Research Foundation under grant No. 140.
  
This work made use of data stored at Barbara A. Mikulski Archive for Space Telescopes (MAST), NASA Astrophysics Data System Bibliographic Services and the NASA/IPAC Extragalactic Database (NED). The {\it HST} photometry and spectroscopy used in this paper can be found in MAST: \dataset[10.17909/6g1c-9b66]{http://dx.doi.org/10.17909/6g1c-9b66}.

\vspace{5mm}
\facilities{HST}

\software{\texttt{astropy} \citep{astropy2018,astropy2013},
          \texttt{astroquery} \citep{astroquery},
          \texttt{photutils} \citep{photutils},
          \texttt{matplotlib} \citep{matplotlib},
          \texttt{numpy} \citep{numpy2020},
          \texttt{scipy} \citep{scipy2020},
          \texttt{Source Extractor} \citep{bertin1996}.
          }



\bibliography{bibliography}{}

\end{document}